# IMPROVED PRECISION MEASUREMENT OF THE CASIMIR FORCE

*Anushree Roy, Chiung-Yuan Lin and U. Mohideen**
*Dept. of Physics, Univ. of California, Riverside, CA 92521.*

## ABSTRACT

We report an improved precision measurement of the Casimir force. The force is measured between a large Al coated sphere and flat plate using an Atomic Force Microscope. The primary experimental improvements include the use of smoother metal coatings, reduced noise, lower systematic errors and independent measurement of surface separations. Also the complete dielectric spectrum of the metal is used in the theory. The average statistical precision remains at the same 1% of the forces measured at the closest separation.

Casimir calculated an extraordinary property that two uncharged metallic plates would have an attractive force in vacuum [1,2]. The force results from the alteration by the metal boundaries of the zero point electromagnetic energy $E = \sum_{n}^{\infty}(1/2)\hbar\omega_n$, where $\hbar\omega_n$ is the photon energy in each allowed photon mode $n$. Lifshitz [3] generalized the force to any two infinite dielectric half-spaces as the force between fluctuating dipoles induced by the zero point electromagnetic fields and obtained the same result as Casimir for two perfectly reflecting (infinite conductivity) flat plates. The Casimir force has been demonstrated between two flat plates [4] and a large sphere and a flat plate [5,6] and its value shown to be in agreement with the theory to an average deviation of 1% [7,8]. For dielectric bodies the resulting force has been measured with reasonable agreement to the theory[9]. Theoretical treatments of the Casimir force have shown that it is a strong function of the boundary geometry and spectrum [10-12]. Experiments with periodically corrugated boundaries have also demonstrated the nontrivial boundary dependence of the Casimir force [13]. Also Casimir force measurements place strong limits on hypothetical long range forces and light elementary particles such as those predicted by supersymmetric theories [14,15]. Thus continuous improvements in the experimental measurement is necessary in order to understand the exact nature of the vacuum fluctuation forces and the interaction of the zero point electromagnetic field with materials. Here we report an improved precision measurement of the Casimir force between a metallized sphere of diameter 201.7 μm and a flat plate using an Atomic Force Microscope (AFM). The results being reported here are an improved version of our earlier experiment[7]. The particular experimental improvements are (i) use of smoother metal coatings, which reduces the effect of surface roughness and allows for closer separations between the two surfaces (ii) vibration isolation which reduces the total noise (iii) independent electrostatic measurement of the surface separations and (iv) reductions in the systematic errors due to the residual electrostatic force, scattered light and instrumental drift. Also the complete dielectric properties of Al is used in the theory. The average precision defined on the rms deviation between experiment and theory remains at the same 1% of the forces measured at the closest separation. The measurement is consistent with the theoretical corrections calculated to date.

The Casimir force for two perfectly conducting parallel plates of unit area separated by a distance $z$ is: $F(z)=-\frac{\pi^2\hbar c}{240}\frac{1}{z^4}$. It is strong function of '$z$' and is measurable only for $z$ < 1 μm. Experimentally it is hard to configure two parallel plates uniformly separated by distances less than a micron. So the preference is to replace one of the plates by a metal sphere of radius $R$ where $R$>>$z$. For such a geometry the Casimir force is modified to [16]: $F_c^0(z)=\frac{-\pi^3}{360}R\frac{\hbar c}{z^3}$. This definition of the Casimir force holds only for hypothetical metals of infinite conductivity, and therefore a correction due to the finite conductivity of Al has to be applied. Such a correction can be accomplished through use of the Lifshitz theory [3,17]. For a metal with a dielectric constant $\varepsilon$ the force between a large sphere and flat plate is given by[3]:

$$F^0(z)=-\frac{R\hbar}{\pi c^3}\int dz\int_0^\infty\int_1^\infty p^2\xi^3 dp\,d\xi\left\{\left[\frac{(s+p)^2}{(s-p)^2}e^{\frac{2p\xi z}{c}}-1\right]^{-1}+\left[\frac{(s+p\varepsilon)^2}{(s-p\varepsilon)^2}e^{\frac{2p\xi z}{c}}-1\right]^{-1}\right\} \qquad (1)$$

where 'z' is the surface separation, $R$ is the sphere radius, $s=\sqrt{\varepsilon-1+p^2}$, $\varepsilon(i\xi)=1+\frac{2}{\pi}\int_0^\infty \frac{\omega\varepsilon''(\omega)}{\omega^2+\xi^2}d\omega$ is the dielectric constant of Al and $\varepsilon''$ is its imaginary component [3]. $\xi$ is the imaginary frequency given by $\omega=i\xi$. Here the complete $\varepsilon''$ extending from 0.04eV to 1000eV from Ref. [18] along with the Drude model below 0.04eV is used to calculate $\varepsilon(i\xi)$. Al metal was chosen because of its high reflectivity at short wavelengths (corresponding to the close surface separations). Alternatively, as Al is a simple metal and its $\varepsilon$ can be well represented with the free electron Drude model of metals with a plasma frequency $\omega_p$[18]. In such a representation $\varepsilon(i\xi)=1+\frac{\omega_p^2}{\xi^2+\gamma\xi}$ is the dielectric constant of Al, $\xi$ is the imaginary frequency given by $\omega=i\xi$, $\omega_p$ is the plasma frequency corresponding to a wavelength of 100nm and $\gamma$ is the relaxation frequency corresponding to 63meV [18]. The two representations of $\varepsilon(i\xi)$, lead to results differing by a few percent for surface separations and experimental uncertainties being reported.

There are also corrections to the Casimir force resulting from the roughness of the surface due to the stochastic changes in the surface separation[7,19]. The roughness of the metal surface is measured directly with the AFM. The metal surface is composed of separate crystals on a smooth background. The height of the highest distortions are 14nm and intermediate ones of 7nm both on a stochastic background of height 2nm with a fractional surface areas of 0.05, 0.11 and 0.84 respectively. The crystals are modeled as parallelepipeds. This leads to the complete Casimir force including roughness correction given by: $F^r(z) = F^0(z)\left[1 + 0.86\left(\frac{A}{z}\right)^2 + 1.02\left(\frac{A}{z}\right)^3 + 1.9\left(\frac{A}{z}\right)^4\right]$ [8]. Here, $A$=11.8nm is the effective height defined by requiring that the mean of the function describing the total roughness is zero and the numerical coefficients are the probabilities of different distance values between the interacting surfaces [8]. The roughness correction here is ≤1.3% of the measured force. There are also corrections due to the finite temperature [20] given by:

$$F_c(z) = F^r(z)\left(1 + \frac{720}{\pi^2} f(\eta)\right), \tag{3}$$

where $f(\eta) = (\eta^3/2\pi)\zeta(3) - (\eta^4\pi^2/45)$, $\eta = 2\pi k_B Tz/hc = 0.131\times10^{-3}z$ $\mathrm{nm}^{-1}$ for T= 300°K, ζ(3)=1.202… is the Riemann zeta function and $k_B$ is the Boltzmann constant. The temperature corrections are less than 1% of the Casimir force for the surface separations reported here.

A schematic diagram of the experiment is shown in figure 1. The force is measured at a pressure below 50mTorr and at room temperature. As in the previous version the experiments were done on a floating optical table. Additionally the vacuum system was mechanically damped and isolated to decrease the vibrations coupled to the AFM. Polystyrene spheres were mounted on the tip of 320µm long cantilevers with Ag epoxy. A 1 cm diameter optically polished sapphire disk is used as the plate. The cantilever (with sphere) and plate were then coated with 250nm of Al in an evaporator. Both surfaces are then sputter coated with 7.9±0.1nm layer of 60%Au/40%Pd (Measured >90% transparency for λ<300nm [7,18] and is thus a transparent spacer for surface separations being considered here. The thickness and the transparency of the Au/Pd sputter coating was measured on a similarly coated glass plate.). This coating was necessary to provide a non-reactive surface and to prevent any space charge effects due to patch oxidation of the Al coating. The sphere diameter was measured using the Scanning Electron Microscope (SEM) to be 201.7±0.5μm. The rms roughness amplitude of the Al surfaces was measured using an AFM to be 3nm. The decrease in roughness was achieved with controlled metal evaporation and reduced coating thickness. The roughness of the metal coating prevents *a priori* knowledge of the average separation on contact of the two surfaces. In the previous experiment [7] this surface separation on contact was estimated from the measured force at large separation distance. Here an independent and exact measurement of the average surface separation on contact of the two surfaces is done by electrostatic means.

In the schematic shown in figure 1, a force on the sphere causes the cantilever to tilt. This tilt is detected by the deflection of the laser beam leading to a difference signal between photodiodes A and B. This force and the corresponding cantilever deflection are related by Hooke's Law: $F=-k\,\Delta z$, where 'k' is the force constant and '$\Delta z$' is the cantilever deflection. As reported in ref.7 the cantilever is calibrated and the residual potential difference between the grounded sphere and plate is measured using the electrostatic force between them. The electrostatic force between the large sphere and the flat surface is given by[21]:

$$F = 2\pi\varepsilon_o (V_1 - V_2)^2 \sum_{n=1}^{\infty} \operatorname{csch} n\alpha \left(\coth\alpha - n\coth n\alpha\right) \quad . \qquad (4)$$

Here '$V_1$' and '$V_2$' are voltages on the flat plate and sphere respectively. $\alpha=\cosh^{-1}\left(1+\frac{z+z_0}{R}\right)$, where $R$ is the radius of the sphere, $z$ is distance between the surfaces, measured from contact and $z_o$ is the true average separation on contact of the two surfaces due to the stochastic roughness of the Aluminum coating. The force constant was measured as k =0.0169±0.0003N/m from the electrostatic force for surface separation > 2μm as reported in ref. [7]. Next the residual potential of the grounded sphere was measured as $V_2$=7.9±0.8mV by the AC measurement technique reported earlier [7] (factor of 3.5 improvement over ref. [7]). This residual potential is a contact potential that arises from the different materials used to fabricate the sphere and the flat plate. The corrections due to the piezo hysteresis and cantilever deflection were applied as reported in ref.[7] to the sphere-plate separations in all collected data

To measure the Casimir force between the sphere and flat plate they are both grounded together with the AFM. The plate is then moved towards the sphere and the corresponding photodiode difference signal was measured (approach curve). The raw data from one scan is shown in Fig. 2. Region-1 is the flexing of the cantilever resulting from the continued extension of the piezo after contact of the two surfaces. In region-2 ($z_0$+516nm>surface separations>$z_o$+16nm) the Casimir force is the dominant characteristic far exceeding all systematic errors. The systematic effects are primarily from the residual electrostatic force (<1.5% of the force at closest separation) and a linear contribution from scattered light. This linear contribution due to scattered light (and some experimental drift) can be observed and measured in region 3.

Next we use of the electrostatic force between the sphere and flat plate to arrive at an independent and consistent measurement of $z_{o}$, the average surface separation on contact of the two surfaces. This is done immediately following the Casimir force measurement without breaking the vacuum and no lateral movement of the surfaces. The flat plate is connected to a DC voltage supply while the sphere remains grounded. The applied voltage $V_1$ in eq. 4 is so chosen that the electrostatic force is >10 times the Casimir force. The open squares in figure 3 represent the measured total force for an applied voltage of 0.31 V as a function of distance. The force results from a sum of the electrostatic force and the Casimir force of eq. 3. The solid line which is a best $\chi^2$ fit for the data in figure 3 results in a $z_o$=47.5nm. The approximation to the electrostatic force given by [13,16] $F_e = \frac{-\pi\varepsilon_0 R}{z + z_o}(V_1 - V_2)^2$ is used in the fit. The experiment is repeated for other voltages between 0.3-0.8 V leading to an average value of $z_o$=48.9±0.6nm (the rms deviation is 3nm). Given the 7.9nm Au/Pd coating on each surface this would correspond to a average surface separation 48.9±0.6+15.8= 64.7±0.6nm for the case of the Casimir force measurement.

The electrostatically determined value of $z_o$ can now be used to apply the systematic error corrections to the force curve of figure 2. Here the force curve in region-3, is fit to a function: $F = F_c(z+64.7\text{nm}) + F_e(z+48.9\text{nm}) + Cz$ . The first term is the Casimir force contribution to the total force in region 3. The second term represents the electrostatic force between the sphere and flat plate due to the residual potential difference of $V_2$=7.9mV. The third term C represents the linear coupling of scattered light from the moving plate into the diodes and experimental drift and corresponds to a force <1pN (<1% of the forces at closest separation). The difference in $z_o$ in the electrostatic term and the Casimir force is due to the 7.9nm Au/Pd coating on each surface. The value of $C$ is determined by minimizing the $\chi^2$. The value of $C$ determined in region 3 and the electrostatic force corresponding to $V_2$=7.9mV and $V_1$=0 is used to subtract the systematic errors from the force curve in region-3 and 2 to obtain the measured Casimir force as: $F_{c-m}= F_m-F_e- Cz$ where $F_m$ is the measured total force. Thus the measured Casimir force from region 2 has no adjustable parameters.

The experiment is repeated for 27 scans and the average Casimir force measured is shown as open squares in figure 4. The error bars represent the standard deviation from the 27 scans at each data point. Due to the surface roughness, the averaging procedure introduces ±3nm uncertainty in the surface separation on contact of the two surfaces. The theoretical curve given by eq.3 with values of $\varepsilon''$ from ref. [18] is shown as a solid line. The theory has no adjustable parameters.

A variety of statistical measures can be used to define the precision of the Casimir force measurement. A key point to note is that the Casimir force is generated for the whole range of separations and is compared to the theory with *no adjustable parameters*. Here we restrict the measurement to surface separations corresponding to wavelengths where Al can be considered a highly reflective metal and the Au/Pd cap layer is largely transparent. Thus we check the accuracy of the theoretical curve over the complete region between 100-500nm with $N$=441 points (with an average of 27 measurements representing each point) with no adjustable parameters. Given that the experimental standard deviation over this range is 7pN from thermal noise, the experimental uncertainty is $\leq \frac{7}{\sqrt{27}} = 1.3pN$ leading to a precision which is better than 1% of the largest forces measured. If one wished to consider the rms deviation of the experiment ($F_{experiment}$) from the theory ($F_{theory}$) in eq.3, $\sigma = \sqrt{\frac{(F_{theory} - F_{\exp eriment})^2}{N}}$ =2.0pN as a measure of the precision, it is also on the order of 1% of the forces measured at the closest separation. Another measure of the precision is through the over 99% confidence level which is obtained from the $N$=441 independent comparisons to theory and a reduced $\chi^2$=0.9. Thus by any of the above definitions, the statistical measure of the experimental precision is of order 1% of the forces at the closest separation.

As regards the effect of the experimental uncertainties, the error from the measurement of the surface separation distance $z$ dominates due to the ~$z^{-3}$ dependence of the Casimir force between the sphere and the flat plate in this region. If the uncertainty of -3nm and +3nm in the measurement of the surface separation is taken as the largest measure of this uncertainty, then $\sigma$ changes to 3.0pN and 3.6pN respectively. To place a more drastic limit, if both 7.9nm Au/Pd cap layers are not transparent as assumed, then the separation on contact is 49.8nm leading to a $\sigma$ of 12.3pN.

In conclusion, we have performed an improved precision measurement of the Casimir force between a large Al coated sphere and flat plate. The experimental improvements over the previous measurement are: (i) use of smoother metal coatings, which reduces the effect of surface roughness and allows for closer separations between the two surfaces, (ii) vibration isolation which reduces the total noise, (iii) independent electrostatic measurement of the surface separations and, (iv) reductions in the systematic errors due to the residual electrostatic force, scattered light and instrumental drift. Also the complete dielectric properties of Al is used in the theory. All the above improvements allow unambiguous comparison of experiment and theory with no adjustable parameters. The average precision remains at the same 1% of the forces measured at the closest separation. The measurement is consistent with theoretical corrections calculated to date.

Discussions with G.L. Klimchitskaya, and V.M. Mostepanenko are acknowledged.

* To whom correspondence should be addressed. Email address: umar.mohideen@ucr.edu

## FIGURE CAPTIONS

Figure 1: Schematic diagram of the experimental setup. Application of voltage to the piezo results in the movement of plate towards the sphere.

Figure 2: A typical force curve as a function of the distance moved by the plate.

Figure 3: The measured electrostatic force for a applied voltage of 0.31 V to the plate. The best fit solid line shown leads to a $z_o$=47.5nm. The average of many voltages leads to $z_o$=48.9±0.6nm.

Figure 4: The measured average Casimir force as a function of plate-sphere separation is shown as squares. The error bars represent the standard deviation from 27 scans. The solid line is the theoretical Casimir force from eq. 3 and tabulated values of $\varepsilon$" with no adjustable parameters.

Fig. 1 Roy, Lin, & Mohideen

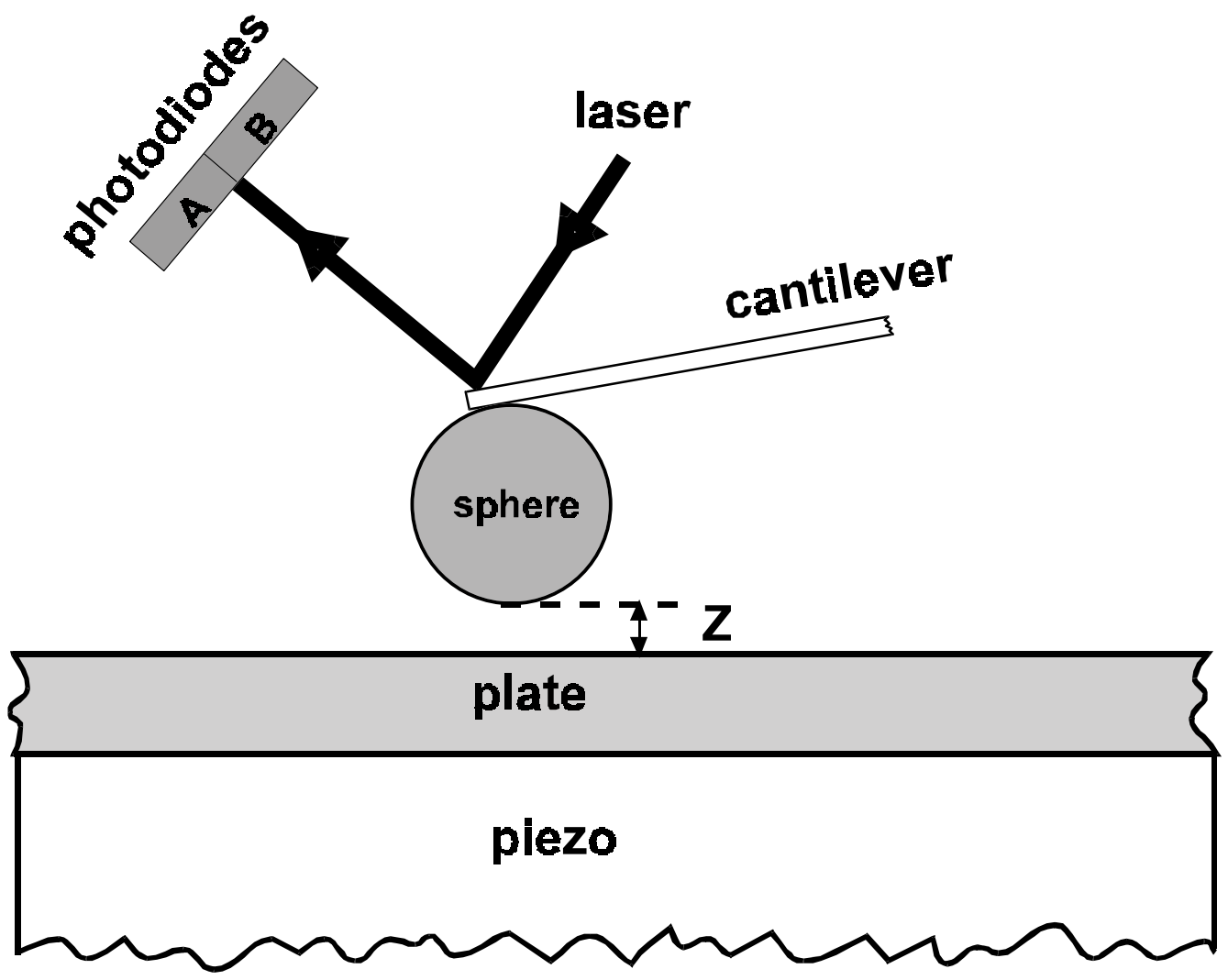

Fig. 2 Roy, Lin, & Mohideen

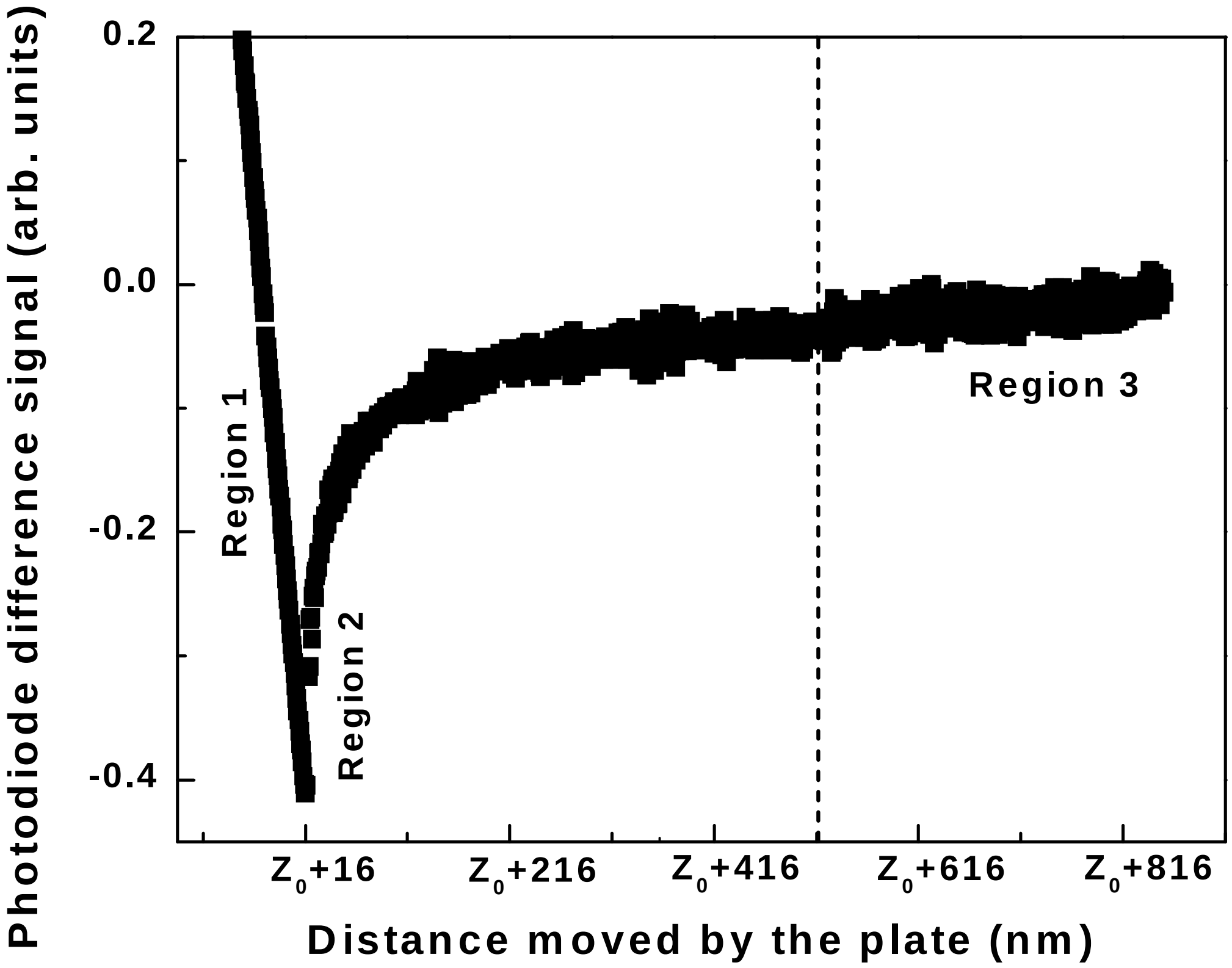

Fig. 3 Roy, Lin, & Mohideen

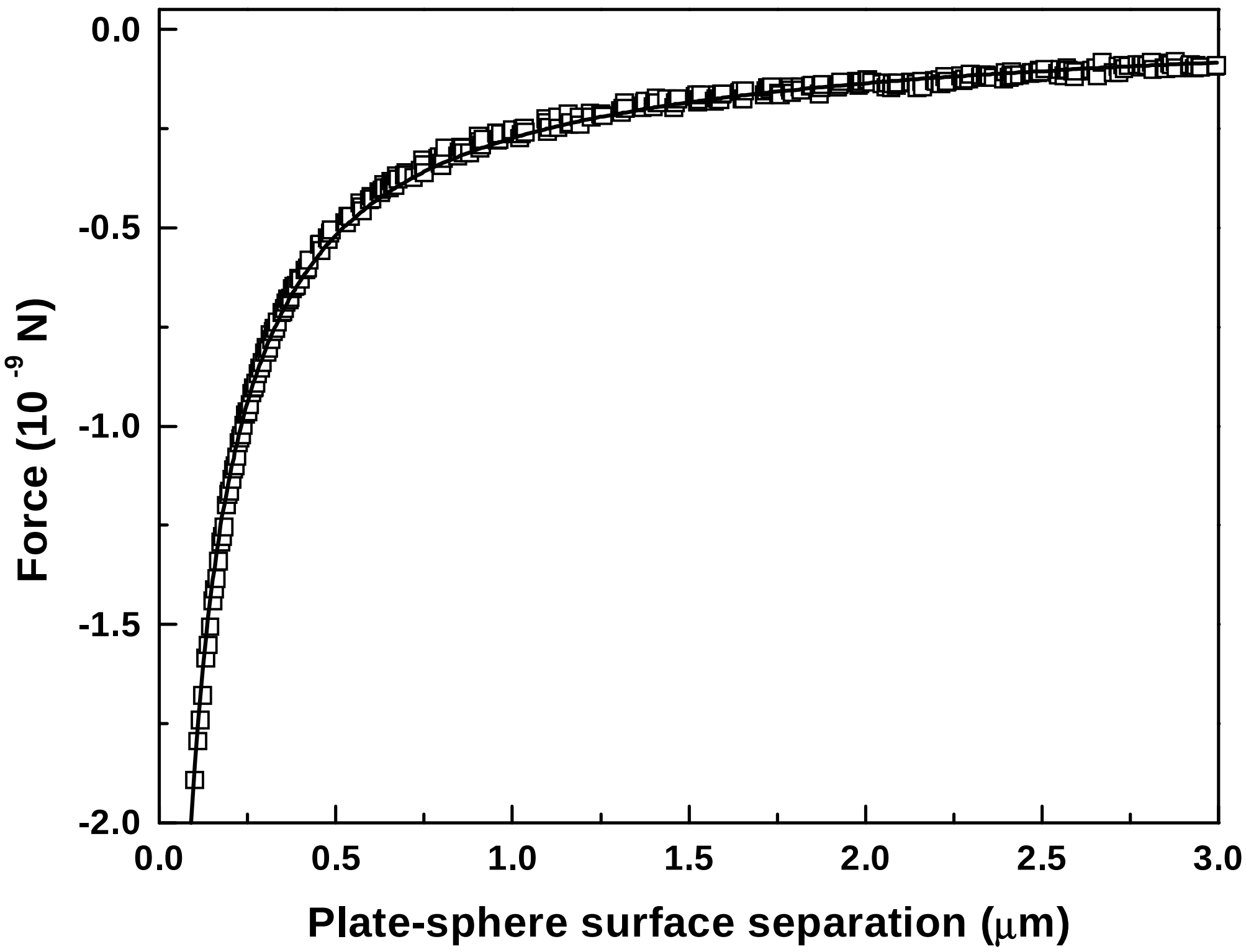

Fig. 4 Roy, Lin, & Mohideen

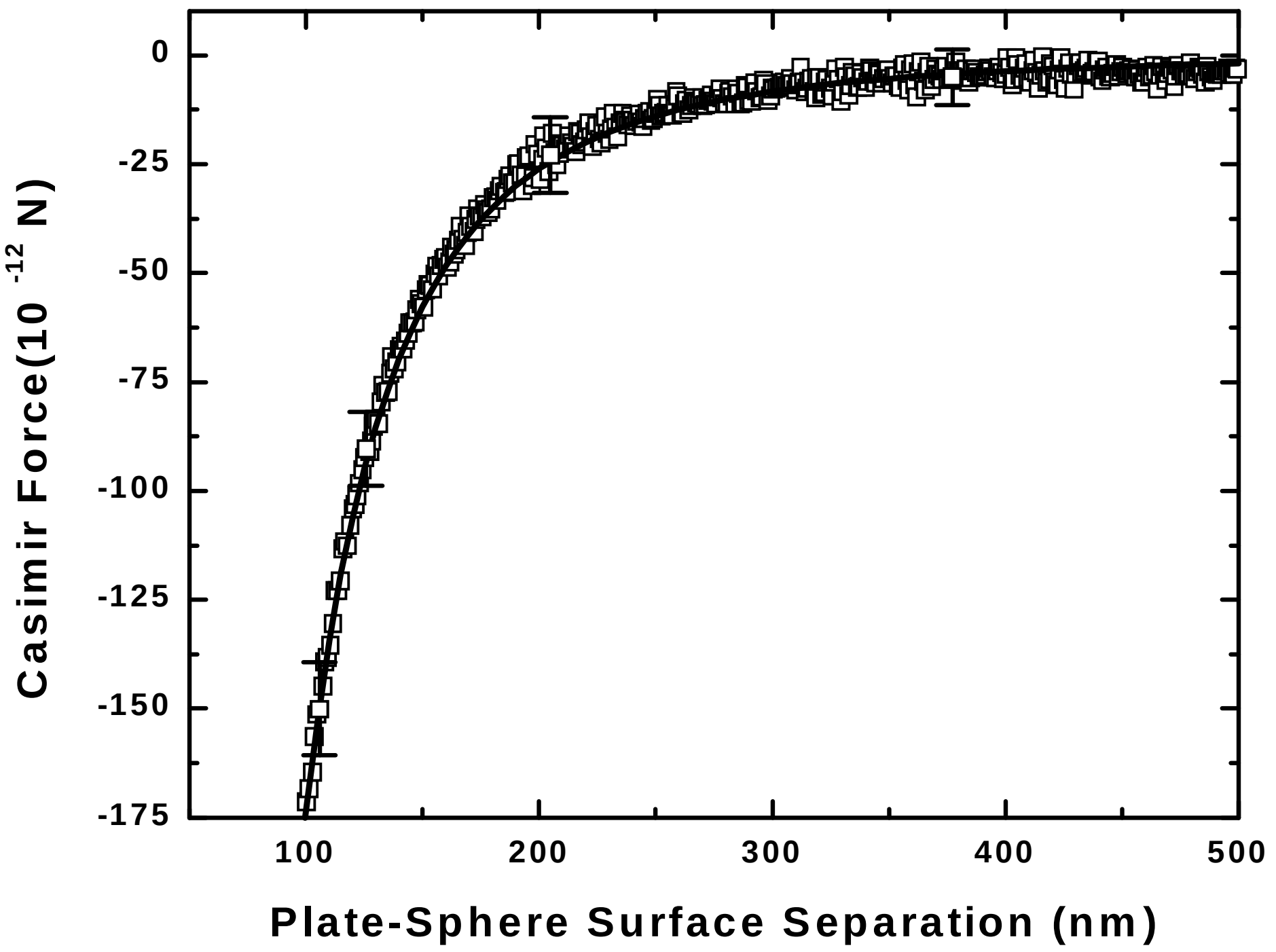